\documentclass[aps,twocolumn,showpacs,floatfix,a4paper]{revtex4}

\usepackage{graphicx}
\usepackage{color}
\usepackage{dcolumn}
\usepackage{amsmath}
\usepackage{amsfonts}
\usepackage{amssymb}

\begin{document}

\bibliographystyle{apsrev}
\title{Quantum Correlations with a Classical Apparatus}
\author{Frederick H. Willeboordse}

\affiliation{
    Dept of Physics, The National University of Singapore, Singapore 119260
 }
\email{willeboordse@yahoo.com}
\homepage[\\Homepage:  ]{http://www.willeboordse.ch/science/}

\begin{abstract}
A deterministic, relativistically local  and thus classical Bell-type apparatus is reported that violates the Bell-CHSH inequality by introducing a simple local memory element in the detector and by requiring the detector combinations to switch with unequal probabilities. This indicates that the common notion of the fundamental impossibility of a classical-type theory underlying quantum mechanics may need to be re-evaluated.
\end{abstract}

\maketitle

\section{Introduction}

A cornerstone in the understanding of the fundamental nature of quantum mechanics is formed by the Clauser-Horne-Shimony-Holt (CHSH) inequality \cite{CHSH} which is based on John S. Bell's reexamination \cite{Bell_Physics_1} of the Einstein-Podolsky-Rosen paradox \cite{EPR}. The CHSH inequality concerns  possible (anti-) correlations in an experimental setup where a source emits an entangled pair of particles that are then independently detected with a measuring device which has (at least) two possible settings and which has a binary output.  Denoting the settings of the first measuring apparatus as $A_1$ and $A_2$, the settings of the second measuring apparatus as $B_1$ and $B_2$, and the probability that the binary outputs of settings $A_i$ and $B_j$ ($i,j = 1,2$) mismatch as $P(A_i,B_j)$, CHSH showed that C = $P(A_1,B_1) + P(A_1,B_2) +  P(A_2,B_1) - P(A_2,B_2)  \leq 2$ for what are generally considered to be \textit{all} possible classical relativistically local binary valued systems. Under the same conditions, quantum mechanics predicts the upper bound to be  $ 2 \sqrt{2}$ instead of  $2$. Experiments by Aspect and others subsequently clearly showed the quantum mechanical value to be correct \cite{Aspect_PRL-49}.

The general consensus that arose from the discrepancy between the experimental results and the limits imposed on classical correlations by the   CHSH inequality is that it is impossible to construct a classical type theory that can fully account for the observed (anti-) correlations \cite{Mermin_PhysicsToday1985,Hess-Philipp_PNAS-98,Kaszlikowski_DoctorateThesis,Shimony_Stanford}. Consequently, it is believed that quantum mechanics cannot be viewed as the statistical mechanics of an underlying local realistic sub-atomic world and any attempt at describing nature by means of e.g. a cellular automaton or a complex system is futile. 

Therefore, first and foremost one needs to get past the hurdle of the CHSH inequality if one would like to open the door to re-evaluating the ideas of realistic descriptions. From a conceptual point of view, getting past this hurdle does not require the actual design of an apparatus or automaton that represents a real physical system as falsification of the consensus interpretation can be achieved by a single counter example as long as it reasonably can be argued that it is genuinly classical. 

This paper is organized as follows: Section \ref{sec:apparatus} describes the setup of the apparatus and its components. Section \ref{sec:numexp} gives details on the numerical experiment as well as the internal workings of the components described in section \ref{sec:apparatus}. For the apparatus, the distinction between weak and strong locality is important and its relevance outlined in section \ref{sec:weakstrong}. A discussion and the conclusions are presented in section \ref{sec:discussion}.

\section{The Apparatus - Classical Model \label{sec:apparatus}}

\begin{figure}[htb]
   \includegraphics*[width=8cm]{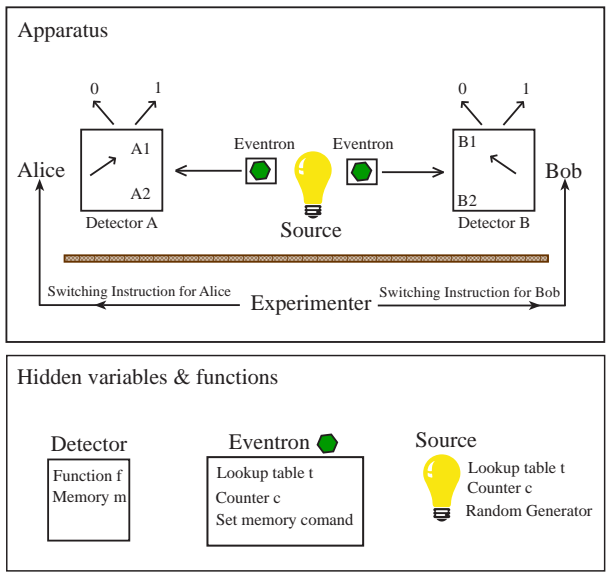}
   \caption
   {
      The apparatus employed. A source in possession of a lookup table and a counter emits two identical eventrons that are processed independently by detectors $A$ and $B$. The experimenter who determines the detector settings does not have any knowledge of the values of the variables carried by the eventron or the detector output.
   }
   \label{apparatus}
\end{figure}

The setup numerically simulated below is depicted in Fig.~\ref{apparatus}. It consists of a source $S$ which periodically emits two identical  ``eventrons" in opposite directions and two detectors $A$,$B$ each with two settings denoted by the subscripts $1$ and $2$ respectively. The detectors are operated by Alice and Bob who receive instructions on how to set the detectors (i.e. what setting the detectors should be in) from the experimenter. When an eventron hits a detector, the detector will output either a one or a zero creating an event that is recorded by Alice and Bob together with the detector setting. All events are recorded (the dropping of events is not permitted). Alice and Bob are not allowed to communicate with each other, back to the experimenter or the source and consequently Alice is not aware of the outcomes recorded by Bob and vice versa. Furthermore, Alice and Bob do not know each others instructions (i.e. the experimenter does not send the instructions for Bob to Alice or vice versa: Alice and Bob only receive their own instructions), nor are they allowed to have predetermined knowledge of each others settings (by means of a published switching schedule e.g.).  

The source is in possession of a binary lookup table $t$ and has a memory-like variable $c$ that acts as a counter and index to entries in the lookup table. The variable $c$ is increased by one every time a pair of eventrons is emitted. The lookup table is created by the source by randomly setting its entries to zero or one with a probability $p_t$ of successive entries being different and without further arrangement of the entries in groups. It is created once at the beginning of an experimental run and is then used throughout the run. It has a given length, though the value is not essential here, and an index $k$ can be used to access entry $t[k]$ (hence the table is like an array in the context of computer programming). When an index $k$ is incremented and subsequently exceeds the table length, it is set to point to item 2 in the table as the entry $k-1$ is used in the computation of the output in some cases. 

The detectors contain local memory-like variables $m_X$  that act as indices to the lookup table, and local functions $f(t,X_i,m_X,c)$ with $X$ indicating either detector $A$ or $B$ and $i$ either setting $1$ or $2$ that determine the binary output based on the value of this internal memory-like variable and  the variables carried by the eventron.  

The eventrons carry with them the source's lookup table, the value of the source's counter $c$ as well as a set-memory command which the source sets with a  probability $p_s$ labeling it randomly as destined for either detector $A$ or detector $B$. When a detector receives a set-memory command destined for it, the variable $m_X$ is set to $m_X = c$.

An experimental run contains the following parts:
\begin{enumerate}
\item Switch on: The source creates the binary lookup table. All the variables of all the components of the apparatus are set to random values, as are the detector settings.
\item During an experimental run: The source periodically emits eventrons. Alice and Bob set the detectors according to the instructions received by the experimenter. They record the detector output and its setting for each event.
\item Switch off: Alice and Bob send the recorded detector outputs and settings to the experimenter who calculates the correlations.
\end{enumerate}
Note that in this setup, it is not necessary to directly synchronize the experimenters' instructions with the source. As none of the eventrons is dropped, he only needs to specify the sequential number of eventrons per setting (e.g. the first four eventrons are measured in setting 1, the next 9 eventrons in setting 2, and so on). However, it would be straightforward to modify the apparatus to allow for dropped eventrons by including a timing mechanism.

Besides the memory-like variables in the detectors, a key difference with conventional setups is that experimenter sends random detector-setting instructions to Alice and Bob such that certain detector setting combinations occur more frequently than others. However it should be stressed that the experimenter does not know outcomes at any stage during a run, and it is argued in the discussion that this neither violates relativistic locality not that it is 'un'-classical.

\section{Details of the Numerical Experiment \label{sec:numexp}}

During one experimental run, the following steps are repeatedly carried out:
 \newline
 
 \noindent \textbf{At the source} 
      \begin{enumerate}
         \item Increase $c$ by one (if it exceeds the table length, set it to point to item 2 of the table)
	 \item Create two identical eventrons
	 \item Copy the value of $c$ to the eventrons
	 \item Copy the lookup table $t$ to the eventrons
	 \item With probability $p_s$ set the set-memory command of the eventrons to either $A_1$ or $B_1$ (randomly chosen). Otherwise, set the command to inactive.
	 \item Launch the eventrons towards the detectors
     \end{enumerate} 
     
\noindent \textbf{During flight of the eventrons}
\newline

	 With probability $p_d$, the experimenter instructs Alice and Bob to set the detectors to $A_i$ and $B_j$ respectively with $i$ and $j$ randomly chosen when the previous detector pairing (only known to the experimenter) was either $A_1$,$B_2$ or $A_2$,$B_1$, while he does so  with a probability $\alpha p_s$, when the previous detector pairing was either $A_1$,$B_1$ or $A_2$,$B_2$ . 
	 The parameter $\alpha$ is a factor ranging from roughly 0.7 to $1/p_s$ (here $\alpha =2$ was used). As $p_d$ and $p_s$ are quite a bit smaller than one, for the majority of events, the detector settings do not change. 
	 Although in the numerical simulations conducted, the experimenter's instructions are dispatched during the flight of the eventrons, this is not essential. Given the probabilities $p_d$ and $p_s$, he could make switching list beforehand. It is only essential that he is ignorant of the detectors' outputs and the values of the sources' variables, and of course that the list is not distributed to the source.

   \vspace*{5mm}    

\noindent \textbf{At the detectors} 
\newline

\noindent  When an eventron arrives at the detector, carry out the function corresponding to the detector setting: \vspace{1cm}

\noindent $f(t,X_1,m_X,c$):
     \begin{itemize}
        \item 
	If the set-memory command indicates $X$, set $m_X = c $. Otherwise leave $m_X$ as it is.
	\item Lookup the entry $t[m_X]$ and emit this entry.
	\item Increase the value of $m_X$ by 1 (if it exceeds the table length, set it to point to item 2 of the table).
     \end{itemize} 
  
\noindent $f(t,X_2,m_X,c$):
    \begin{itemize}
       \item  Lookup the entry $t[c-1]$ and emit this entry
    \end{itemize}

\noindent
The apparatus can violate the CHSH inequality due to a combination of two factors. 

Firstly, the construction of the lookup table assures that when $c$ and $m_X$ are aligned (i.e. when $c = m_x$), the probability of mismatch between $A_1$ and $B_2$ or between $A_2$ and $B_1$ is equal to $p_t$. As long as $p_s$ is sufficiently larger than $p_d$, this alignment is mostly active for detector pairs $A_1$,$B_2$ or $A_2$,$B_1$.   Detector settings $A_2$ and $B_2$ always yield the same binary output and consequently the probability of mismatch for this combination is zero. 

Secondly, for detector combination $A_1$,$B_1$, the alignment is mostly inactive as the probability that both detectors receive a set-memory command before the detector settings change is rather small. Consequently, for the pairing $A_1$, $B_1$, the respective memories $m_X$ are de-aligned most of the time. As the probability of obtaining a zero or one equals to 1/2 in the lookup table when randomly picking an entry, the probability of mismatch when looking up two values with the two de-aligned indices equals 1/2 as well. Consequently, the lower limit of $p_t$ for violating the Bell-CHSH inequality is given by $2 p_t + 1/2 = 2$. In practice, of course, $A_1$, $B_1$ will be aligned sometimes and hence $p_t$ needs to be sufficiently larger than 0.75 in order to compensate for the lost (anti-)correlations. 

Numerical verification was carried out for $p_t = 0.9$ and a histogram of the results is displayed in Fig. 2. 
\begin{figure}[htb]
   \includegraphics*[width=7cm]{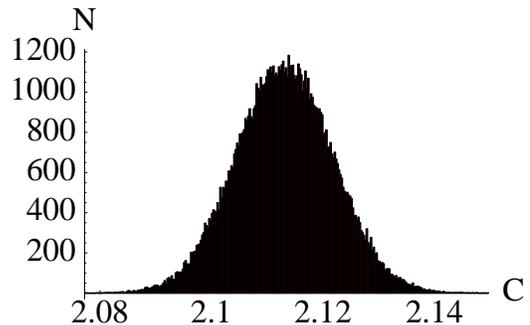}
   \caption
   {
     Histogram for the correlations of $10^5$ runs. For each run, $10^6$ eventrons were emitted from the source. The settings for the numerical experiment were: Table length = 10,000; Probabilities, $p_t = 0.9$, $p_s = 0.1$, $p_d = 0.01$. 
   }
   \label{histogram}
\end{figure}
It should be noted that when considered individually, each detector setting will emit a one with probability 1/2 as can be expected from the construction of the lookup table.

The question that arises immediately is: "What is different so that Bell-CHSH-type proofs do not apply"? 
The essential differences are the inclusion of a memory term and the dependence of the detector output on how long the detector has been in a setting (i.e. a time dependence due to time it takes to receive a set-memory command). These dependences prevent the factorization of the detector probabilities and consequently, Bell's proof is not applicable to the current system. As such, of course, it is not be particularly exciting to find that a system not covered by the conditions of a proof, does not obey that very proof's results. The point, however, is that conclusions (which by themselves naturally reside outside the proof) are drawn which are not warranted, namely the impossibility of a realistic local theory. It may be argued that the inclusion of a memory term is disallowed as subsequent detector readings need to be independent. However, physically speaking, it is hard if not impossible to assess what this would exactly mean if the changes were to occur at levels many orders of magnitude below the observed levels and outcomes were a statistical representation of the changed sub-micro states. For example, the seemingly random output of a single detector is not in conflict with a deterministic underlying process. For practical purposes, even one of the simplest two-state nearest neighbor cellular automatons (rule 30) can provide excellent randomness event though each subsequent state is completely defined by the previous state \cite{Wolfram_ANKOS}.

While the apparatus described in Fig. 1 can easily be simulated numerically as done here, the results apply just as well to a mechanical device of the same design and can therefore experimentally be tested.

\section{Strong and Weak Locality \label{sec:weakstrong}}

Probably the least elegant part of the proposed approach is the requirement that on average the time the apparatus is set to detector combinations $A_1$,$B_1$ must be significantly shorter than the average time that it is set to combinations  $A_1$,$B_2$ or $A_2$,$B_1$ (the total number of events per detector combination over the course of the experiment can however be identical). From a conceptual point of view one can argue however that the settings of the detectors are at the experimenter's discretion. Neither the eventrons nor the detectors are aware of the experimenter's choices while for the apparatus to violate the CHSH inequality it is furthermore not necessary that the experimenter knows the outputs of the detectors or the values of the variables of the source.  Also, successive detector combinations are randomly chosen and not dependent on any of the previous combinations. Therefore, even though as such the timing of the detector setting switches can be considered as responsible for the additional correlations, it does not break relativistic locality and is fully classical. Indeed, this is not non-local either in the sense of Einstein´s principle of separability as described by Aspect \cite{Aspect_PRD-14} in the following manner ``the setting of a measuring device at a certain time (event A) does not influence the result obtained with the other measuring device (event B) if the event B is not in the forward light cone of event A (nor does it influence the way in which particles are emitted by a source if the emission event is not in the forward light cone of event A)". After all, in the described apparatus, there is no communication from detector A to B and if so desired, the detector switching schedule could be generated before the experiment is carried out and dispatched to the detectors in such a way that the source is ignorant of the schedule, and that A and B only know their own schedules but not each others. Hence, it can be ascertained that information remains within the light cones at all times.

In this context it is important to note that Jarrett has shown that the factorization which is an essential condition for Bell-type Theorem proofs can be considered as the conjunction of (relativistic) locality and (statistical) completeness \cite{Jarrett_Nous-18,Ballentine_AmJPhys-55} (see also \cite{Shimony_Stanford}). Consequently, as relativity has extensive experimental support, violation of the Bell theorem implies incompleteness and quantum mechanics is indeed incomplete in this sense (as Jarrett points out this terminology should not mistakingly be taken to imply defectiveness). The crux of the matter, however, is the assertion that deterministic theories always satisfy the completeness condition and that consequently deterministic classical systems are governed by Bell-type theorems \cite{Jarrett_Nous-18,Shimony_Stanford}. As the proposed apparatus shows, this assertion is not necessarily valid for {\it all} classical systems if time and (local) memory are incorporated.

This can also be seen as follows. If the detectors A and B have $\lambda$-dependent memories $\kappa_A (\lambda)$ and $\kappa_B(\lambda)$ respectively, the probability of obtaining outcomes $x_A$ and $x_B$ can be expressed as:
\begin{eqnarray}
   P (  x_A  ,x_B | A,B,\lambda)   
   = \frac{1}{N_{\kappa_A}N_{\kappa_B}} \sum_{\kappa_A''(\lambda)\kappa_B''(\lambda)} P_S \quad \quad \quad \\
                           P_S =              P(x_A,x_B|A(\kappa_A''(\lambda)) B(\kappa_B''(\lambda)),\lambda)
   = \frac{N_{\kappa_A''(\lambda)}N_{\kappa_B''(\lambda)}}{N_{\kappa_A}N_{\kappa_B}} \nonumber
\end{eqnarray}
where the double prime in $\kappa''$ indicates that the sum is to be taken only over those values that yield the outcomes $x_A$ and $x_B$, and $N_v$ the total number of different values a variable $v$ can attain.  As the apparatus is deterministic and relativistically local, the probability $Q_A (Q_B)$  of obtaining $x_A (x_B)$ on the left (right) hand side while ignoring the right (left) hand side is:
\begin{eqnarray}
  Q_A  & = & P(x_A | A, \lambda) = 
  \sum_{\kappa_A'(\lambda)} \sum_{x_B} P(x_A,x_B | A(\kappa_A'(\lambda)),B,\lambda) \nonumber \\
  & = & \frac{N_{\kappa_A'(\lambda)}}{N_{\kappa_A}}  \\
   Q_B  & = & P(x_ B| B, \lambda) =  
   \sum_{\kappa_B'(\lambda)} \sum_{x_A} P(x_A,x_B | A, B(\kappa_B'(\lambda)),\lambda) \nonumber \\
  & =& \frac{N_{\kappa_B'(\lambda)}}{N_{\kappa_B}}
\end{eqnarray}
where the single prime in $\kappa'$ indicates that the sum is taken only over those values that yield the outcome $x_A$. It should be noted that the set of values from $\lambda$ for which $\kappa'(\lambda)$ yields $x_A$ will generally be different from the set of values from $\lambda$ for which both $\kappa_A''(\lambda)$ and $\kappa_B''(\lambda)$ yield outcomes $x_A$ and $x_B$ respectively.
Therfore we obtain,
\begin{eqnarray}
  P (  x_A  ,x_B | A,B,\lambda) & = & \frac{N_{\kappa_A''(\lambda)}N_{\kappa_B''(\lambda)}}{N_{\kappa_A}N_{\kappa_B}} \\
  & \neq & \frac{N_{\kappa_A'(\lambda)}}{N_{\kappa_A}} \frac{N_{\kappa_B'(\lambda)}}{N_{\kappa_B}} = Q_A Q_B
  \nonumber
\end{eqnarray}
showing that factorization is not necessarily possible and thus that strong locality does not hold.

Hence, I believe that the Bell theorem only applies to a subset of all possible classical systems and that completeness  is an additional condition that needs to be justified on grounds other than local realism. This of course does not imply that the completeness condition is not reasonable per se or that it is not an accurate reflection of nature. The only implication is that it needs to be motivated independently from local realism and that hence by itself local realism does preclude quantum correlations. 

\section{Discussion \label{sec:discussion}}

One may nevertheless wonder if, in an indirect way, the proposed experimental apparatus doesn't simply set the detectors based on knowledge about the source's output. One could, e.g., imagine a source periodically emitting instructions (1,0,0,1; 0,1,0,1; 0,0,1,1; 1,0,1,0) corresponding to detectors settings $A_1,A_2,B_1,B_2$ respectively. If the experimenters then would loop through the settings pairs $A_1,B_1 \rightarrow A_1,B_2 \rightarrow A_2,B_1 \rightarrow A_2,B_2$, they would obtain 3 for the (anti-)correlations. However, if the experimenter were one step out of sync, the anti-correlation would be 1. Similarly, for two and three steps out of sync, they would obtain 2. Consequently, for random starting points, inevitable when requiring ignorance of the source's variables, the average over many runs would reduce to the maximal (anti-)correlation of 2. Here, this is not the case, even when executing the experiment many times with different random starting points for each separate part of the experiment (detectors, source, experimenter's switching decisions),  the violation occurs as is shown in Fig. 2. Furthermore, in order to obtain the violation, it is not necessary to carefully tune the value of the probabilities $p_s$ and $p_d$.  The only requirement is that $p_d$ is sufficiently smaller than $p_s$ as is shown in Fig. 3. 
\begin{figure}[htb]
   \includegraphics*[width=6cm]{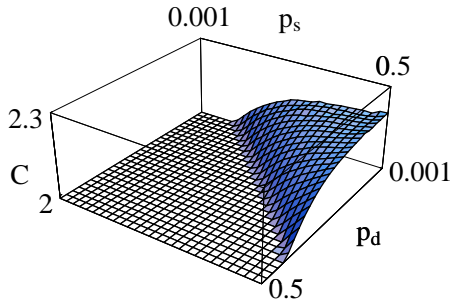}
   \caption
   {
      Log-log plot of detector and set-memory probabilities that lead to correlations larger than 2. The plot shows the correlations obtained for a mesh of 28 values for $p_s$ ranging from 0.001 to 0.5 and 28 values of $p_d$ ranging from 0.0001 to 0.5. For each data point, the average of 100 runs of $10^6$ eventrons was taken. The table length was 10,000 and $p_t = 0.9$. 
   }
   \label{settings}
\end{figure}

The main objective of this report was a proof of concept, namely that contrary to common belief, the CHSH and Bell inequalities do not exclude the possibility of a realistic theory. The proposed apparatus shows that the statistical independence necessary for the factorization condition in the Bell theorem does not directly follow from either (relativistic) locality or determinism and that it hence needs to be justified independently of these concepts. 

It will be interesting to see whether the apparatus can be modified to yield correlations mimicking those found in photon coincidence experiments. The results should provide impetus to efforts of describing nature as a cellular automaton\cite{tHooft_SPIN-2002, Wolfram_ANKOS} or complex system and may have important bearings in the area of secure communication.

\begin{acknowledgments}
I would like to thank Dagomir Kaszlikowski for his extraordinary efforts to explain the intricacies of the Bell theorem to me, Andreas Keil for numerical verification of the results and careful reading of the manuscript, and Michael Revzen, Marek Zukowski, Markus Aspelmeyer and Thomas Osipowicz for fruitful and illuminating discussions.
\end{acknowledgments}

\bibliography{frederik}

\end{document}